\documentclass{aa}
\usepackage{natbib}
\usepackage{subfigure}
\usepackage[varg]{txfonts}

\begin{document}

\title{High-resolution IR absorption spectroscopy of polycyclic aromatic hydrocarbons in the 3 $\mu$m region: role of hydrogenation and alkylation}
\titlerunning{Role of hydrogenation and alkylation in the IR spectroscopy of PAHs}

\author{Elena Maltseva \inst{\ref{inst1}} \and Cameron J. Mackie \inst{\ref{inst2}} \and Alessandra Candian \inst{\ref{inst2}}\and Annemieke Petrignani \inst{\ref{inst2}}\thanks{Current affiliation: University of Amsterdam, Science Park 904, 1098 XH Amsterdam, The Netherlands} \and Xinchuan Huang \inst{\ref{inst3},\ref{inst4}} \and Timothy J. Lee \inst{\ref{inst4}} \and Alexander G. G. M. Tielens \inst{\ref{inst2}} \and Jos Oomens \inst{\ref{inst5}} \and Wybren Jan Buma \inst{\ref{inst1}}}
\institute{University of Amsterdam, Science Park 904, 1098 XH Amsterdam, The Netherlands}

\institute{University of Amsterdam, Science Park 904, 1098 XH Amsterdam, The Netherlands, \email{w.j.buma@uva.nl}\label{inst1}
\and Leiden Observatory, Niels Bohrweg 2, 2333 CA Leiden, The Netherlands \label{inst2}
\and SETI Institute, 189 Bernardo Avenue, Suite 100, Mountain View, CA 94043, USA \label{inst3}
\and NASA Ames Research Center, Moffett Field, California 94035-1000, USA \label{inst4}
\and Radboud University, Toernooiveld 7c, 6525 ED Nijmegen, The Netherlands \label{inst5}
}

\date{Received date / Accepted date }

\abstract {} 
{We aim to elucidate the spectral changes in the 3 $\mu$m region that result from chemical changes in the molecular periphery of polycyclic aromatic hydrocarbons (PAHs) with extra hydrogens (H-PAHs) and methyl groups (Me-PAHs).} 
{Advanced laser spectroscopic techniques combined with mass spectrometry were applied on supersonically cooled 1,2,3,4-tetrahydronaphthalene, 9,10-dihydroanthracene, 9,10-dihydrophenathrene, 1,2,3,6,7,8-hexahydropyrene, 9-methylanthracene, and 9,10-dimethylanthracene, allowing us to record mass-selective and conformationally selective absorption spectra of the aromatic, aliphatic, and alkyl CH-stretches in the 3.175-3.636 $\mu$m region with laser-limited resolution. We compared the experimental absorption spectra with standard harmonic calculations and with second-order vibrational perturbation theory anharmonic calculations that use the SPECTRO program for treating resonances.}
{We show that anharmonicity plays an important$\text{}$ if not dominant role, affecting not only aromatic, but also aliphatic and alkyl CH-stretch vibrations. The experimental high-resolution data lead to the conclusion that the variation in Me- and H-PAHs composition might well account for the observed variations in the 3 $\mu$m emission spectra of carbon-rich and star-forming regions. Our laboratory studies also suggest that heavily hydrogenated PAHs form a significant fraction of the carriers of IR emission in regions in which an anomalously strong 3 $\mu$m plateau is observed.}
{}
        
\keywords{Astrochemistry --- Molecular data --- ISM: molecules --- Infrared: ISM }

\maketitle

\onecolumn

\section{Introduction}

The series of unidentified infrared bands (UIRs) (3.29, 6.2, 7.7, 8.7, and 11.3 $\mu$m) observed in a variety of astrophysical environments has puzzled astronomers for decades since their first discovery in 1973 \citep{Merrill1975,Gillett1973}. The most generally accepted hypothesis is the so-called PAH hypothesis, which asserts that the UIR emission is a result of  radiative cooling of isolated PAHs $\text{}$ (a class of organic compounds consisting of hydrogen and carbon atoms combined in fused aromatic rings)$\text{}$ that have been excited by UV radiation and that populate
the ground state's vibrational manifold after radiationless decay to the ground electronic state  \citep{Sellgren1984,Leger1984,Allamandola1985}. Owing to its accessibility with ground-based telescopes, one of the most frequently studied UIR features is the 3.29 $\mu$m band, which has been attributed to CH-stretch vibrational modes of aromatic hydrocarbons. With higher resolution and sensitivity, observations have revealed that the 3.29 $\mu$m band is accompanied by a plateau spanning the 3.1-3.7 $\mu$m region (the so-called 3 $\mu$m plateau) with superimposed bands at 3.40, 3.46, 3.51, and 3.56 $\mu$m. Observations have also demonstrated that there is a great diversity in the relative strengths of these sub-features depending on the nature of the emission source. In the majority of astronomical environments showing UIRs, the 3.29 $\mu$m band dominates the 3 $\mu$m region \citep{Sloan1997,Diedenhoven2004}. However, there are known sources (mostly protoplanetary nebulae) that possess an abnormal 3 $\mu$m profile in which the plateau and its superimposed bands have the same intensity as the 3.29 $\mu$m band or exceed it in some case \citep{Geballe1992,Geballe1990,Goto2007}.

The identification of the 3 $\mu$m features in combination with the observed variety of profiles is key to a fundamental understanding of carbon evolution in space. It is therefore a subject of extensive discussions. Several suggestions have been put forward to account for the 3.4, 3.46, 3.51, and 3.56 $\mu$m bands. These range from an assignment in terms of hot bands of aromatic CH-stretch transitions $\nu$= 2$\rightarrow$1 and 3$\rightarrow$2 that are redshifted due to anharmonicity \citep{Barker1987a} to invoking alkyl CH-stretch modes in methyl-substituted PAHs (Me-PAHs) \citep{Joblin1996, Pauzat1999,JourdaindeMuizon1990} and aliphatic CH-stretch vibrations of PAHs containing extra hydrogen atoms (H-PAHs) \citep{Bernstein1996,Sandford2013b,Steglich2013a}. Seminal laboratory studies of IR emission in gas-phase UV-excited PAHs \citep{Wagner2000a} did not find indications for the presence of CH-stretch $\nu$= 2$\rightarrow$1 hot-bands that had previously been suggested to account for the 3.4 $\mu$m band \citep{Barker1987a}. It was thus concluded that the satellite 3 $\mu$m features must derive from other carriers. It was also found that the IR emission of H-PAHs is more consistent with astrophysical observations than the IR emission of Me-PAHs \citep{Wagner2000a}. The identification of the carriers of the 3 $\mu$m plateau is still subject of speculation. For example, the previously proposed hypothesis that the plateau is a quasi-continuum of overlapping overtones and combination bands of CC-stretch modes in PAHs \citep{Allamandola1989} found only partial support from our recent studies on linear acenes and condensed PAHs \citep{Maltseva2015,Maltseva2016}. In these studies, experiments and calculations showed unambiguously that aromatic IR activity is restricted to the 3.17-3.33 $\mu$m absorption range, and thus another carrier is required to explain IR activity in the region 3.33-3.64 $\mu$m. Recent extensive matrix-isolation studies and spectroscopic studies on pellets (as grains) \citep{Sandford2013b,Steglich2013a} concluded that CH-stretches in alkyl and methyl groups attached to the periphery of PAHs are the main candidates for the assignment of the weaker bands in the 3 $\mu$m region. However, this claim has thus far not been confirmed because we lack high-resolution gas-phase IR spectra of these species.

In previous studies \citep{Maltseva2016,Maltseva2015}, we applied IR-UV ion-dip laser spectroscopy combined with mass-selective ion detection on PAHs in molecular jets where the seeded molecules could reach internal temperatures lower than 5 K. These conditions allow for the recording of conformational- and mass-selected spectra with laser-limited line widths below 0.1 cm$^{-1}$. We found that the CH-stretch region of regular PAHs is dominated by large Fermi resonance polyads, which result from a plethora of combination bands, and that a high-end treatment of anharmonic effects including resonance polyads is required to characterize this region computationally. We concluded that the shape of the 3 $\mu$m band can act as a spectral signature and help in providing a precise identification of the UIRs carriers. 
    
\begin{figure}[t]
\centering
        \includegraphics[scale=0.5]{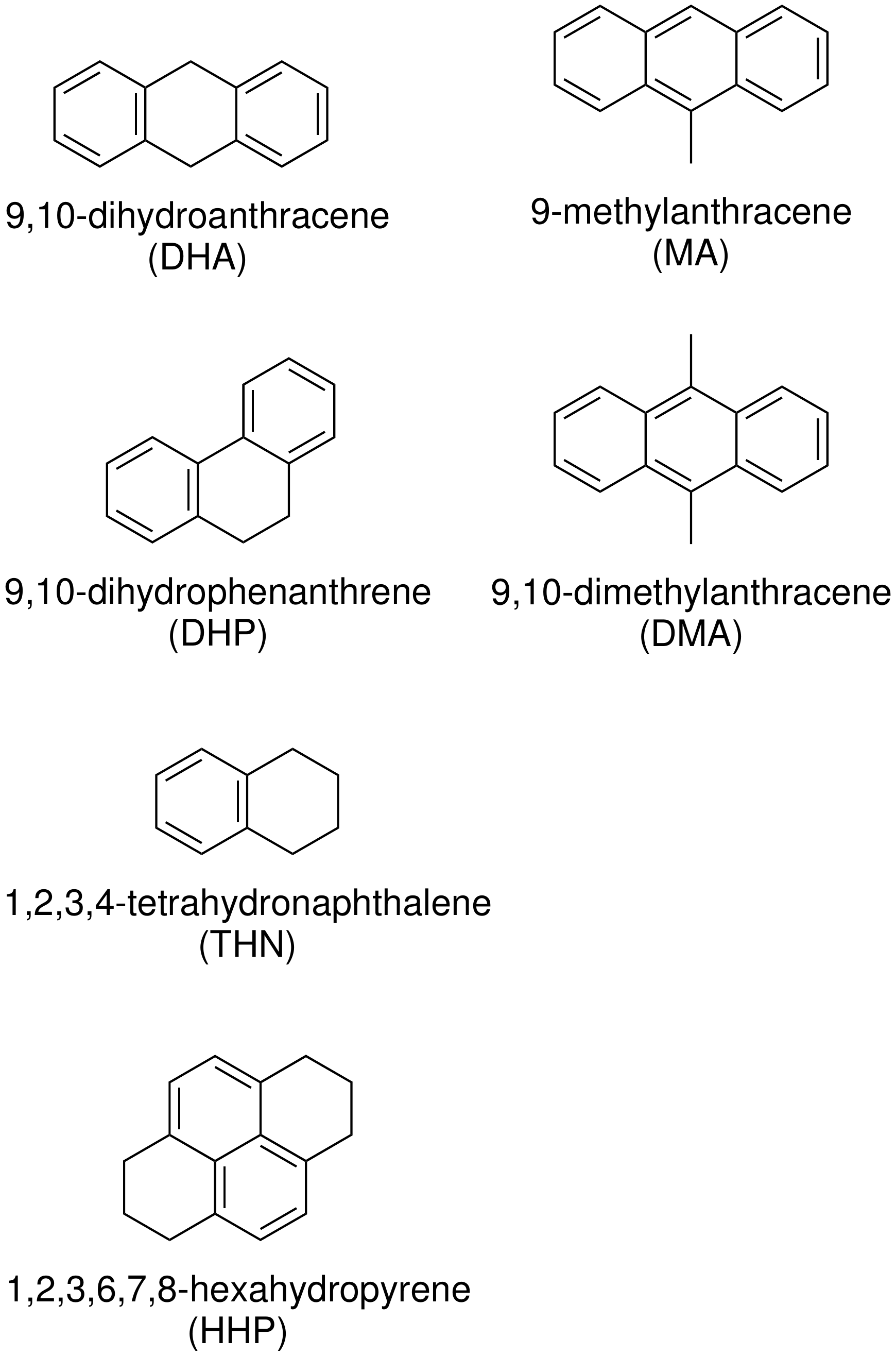}
        \caption{Chemical structures of the hydrogenated PAHs DHA, DHP, THN, and HHP, and of the methylated PAHs MA and DMA. \label{Mol}}
\end{figure}

Similar ``zero''-temperature studies of decorated PAHs are as yet absent. They are, however, much needed as they provide the required benchmark for theoretical studies and thereby for further studies on the emission spectra of highly excited functionalized PAHs. In the present paper we therefore investigate the effects of hydrogenation and methylation on the CH-stretch region using the same approach as we employed in our previous studies on bare PAHs. To this purpose, we have selected four PAHs with an excess of hydrogens (1,2,3,4-tetrahydronaphthalene 
(THN (C$_{10}$H$_{12}$)), 9,10-dihydroanthracene (DHA (C$_{14}$H$_{12}$)), 9,10-dihydrophenathrene (DHP (C$_{14}$H$_{12}$)), 1,2,3,6,7,8-hexahydropyrene (HHP (C$_{16}$H$_{16}$))), and two PAHs with methyl-groups (9-methylanthracene (MA (C$_{15}$H$_{12}$)) and 9,10-dimethylanthracene (DMA (C$_{16}$H$_{14}$))
 (see Fig. \ref{Mol})) and compared their absorption spectra with the spectra of the bare PAH parent compounds that we studied
previously (naphthalene (N), anthracene (A), phenanthrene (Ph), and pyrene (P)). The spectra recorded for the hydrogenated PAHs DHA, DHP, HHP, and THN enable us to determine how hydrogenation of the molecular periphery affects the spectral signatures of the solo, duo, trio, and quarto positions, respectively. At the same time, this set shows how the degree of hydrogenation (from minimally hydrogenated PAHs such as DHA and DHP to heavily hydrogenated PAHs such as HHP and THN) influences the CH-stretch region as a whole and allows us to determine the ratio between aromatic and aliphatic carbon in interstellar space, as was previously done by \cite{Li2012} using the 3.4 $\mu$m and 6.85 $\mu$m emission features. The effects of methylation on the CH-stretch region have been studied by comparing the absorption spectra of bare anthracene with anthracene derivatives that have one (MA) or two (DMA) methyl groups at the solo positions. Aspects that are of particular interest here are whether the aromatic CH-stretch region is affected in the same way as in the parent compound and whether anharmonicity has as large an influence on the aliphatic region as it has on the aromatic region. One further important aspect is that the methyl groups are quasi-free rotors that so far have not been studied with our theoretical method. 
Comparison of the results from the present study at zero-temperature with previously reported high-temperature gas-phase spectra from the NIST catalog \citep{NIST} and with spectra reported in matrix isolation spectroscopic (MIS) studies will allow us to assess the influence of temperature and interactions with an environment on the spectra, and thereby on the conclusions drawn from such spectra in terms of PAH composition and evolution.

With this work we aim first of all to investigate the role of anharmonicity in the alkyl and aliphatic CH-stretch region and determine whether the plateau in the 3.333-3.636 $\mu$m region is primarily due to the effects of anharmonicity. Second, we aim to identify the molecular origins of the 3.4, 3.46, 3.51, and 3.56 $\mu$m UIR emission bands and of the 3 $\mu$m plateau. Finally, we wish to establish spectral fingerprints at a molecular
as well as class-sensitive level for two different classes of decorated PAHs, namely H-PAHs and Me-PAHs. These fingerprints are important since they can help us to shed light on the PAH composition in different astronomical environments.

\section{Methods}

The experiments were carried out using a molecular beam setup described earlier (\cite{Smolarek2011}). In order to obtain cold and isolated molecules, the sample of interest was kept at temperatures above its melting point in the container attached to a pulsed valve (General Valve). Subsequent pulsed expansion in a carrier gas (argon at 2 bars), with a typical pulse duration of 200 $\mu$s, resulted in supersonic cooling of the sample molecules.

Double-resonance UV-IR spectroscopy was used to study the ground-state vibrational manifold of PAHs. A two-color resonance-enhanced multiphoton ionization (REMPI) scheme was used to create an ion signal that was detected in a time-of-flight spectrometer (R. M. Jordan Co.) at the molecular mass.  Importantly, this mass-resolved ion detection approach thus allows us to record IR absorption spectra that are not affected by fragmentation or other possible photoreactions. An excitation laser (Sirah Cobra Stretch) was set to the wavelength of S$_{1}$ $\leftarrow$ S$_{0}$ transitions that are well known from the literature (DHA \citep{Chakraborty1990}, DHP \citep{Chakraborty1991}, THN \citep{Yang2007}, HHP \citep{Chakraborty2001}, MA \citep{Tanaka1986}, and DMA \citep{Hirayama1990}). An excimer ArF laser (193 nm, Neweks PSX-501) was used for the subsequent ionization of the excited molecules. In order to probe IR transitions, an IR laser pulse with a line width of 0.07 cm$^{-1}$ was introduced 200 ns before the ionization and excitation lasers. The 3 $\mu$m beam was produced in a LiNbO$_{3}$ crystal by difference frequency mixing of the fundamental output of a dye laser (Sirah Precision Scan with LDS798 dye) and the 1064 nm fundamental of an Nd:YAG laser (Spectra Physics Lab 190). The spectra were recorded in the range of 3.175-3.636 $\mu$m (3150-2750 cm$^{-1}$). With the current signal-to-noise ratio (S/N), no bands could be observed outside this region.

We compared our experimental spectra with theoretically predicted spectra obtained as described previously \citep{Maltseva2015,Maltseva2016,Mackie2015,Mackie2016},
although the calculations employed the B3LYP \citep{Becke1993, Lee1988} hybrid functional in conjunction with the polarized double-$\zeta$ basis set N07D \citep{Barone2008} \citep{Mackie2017}. We refer to calculations performed at the standard harmonic vibrational level as implemented in Gaussian09 \citep{g09} as G09-h calculations, while anharmonic calculations that use a locally modified version of the program SPECTRO \citep{Gaw1996,Mackie2015,Mackie2015a,Candian2017} are referred to as SP16 calculations. Vibrational frequencies from the harmonic calculations were scaled with a scaling factor (sf) of 0.96, while anharmonic frequencies do not require any scaling. For further comparison with the experiment, both harmonic and anharmonic calculations were convolved with a 4 cm$^{-1}$ Gaussian line shape.

\section{Results}

\subsection{Hydrogenated PAHs}

\begin{figure}[t!]
        \centering      
        \subfigure[DHA]{\includegraphics[scale=0.3]{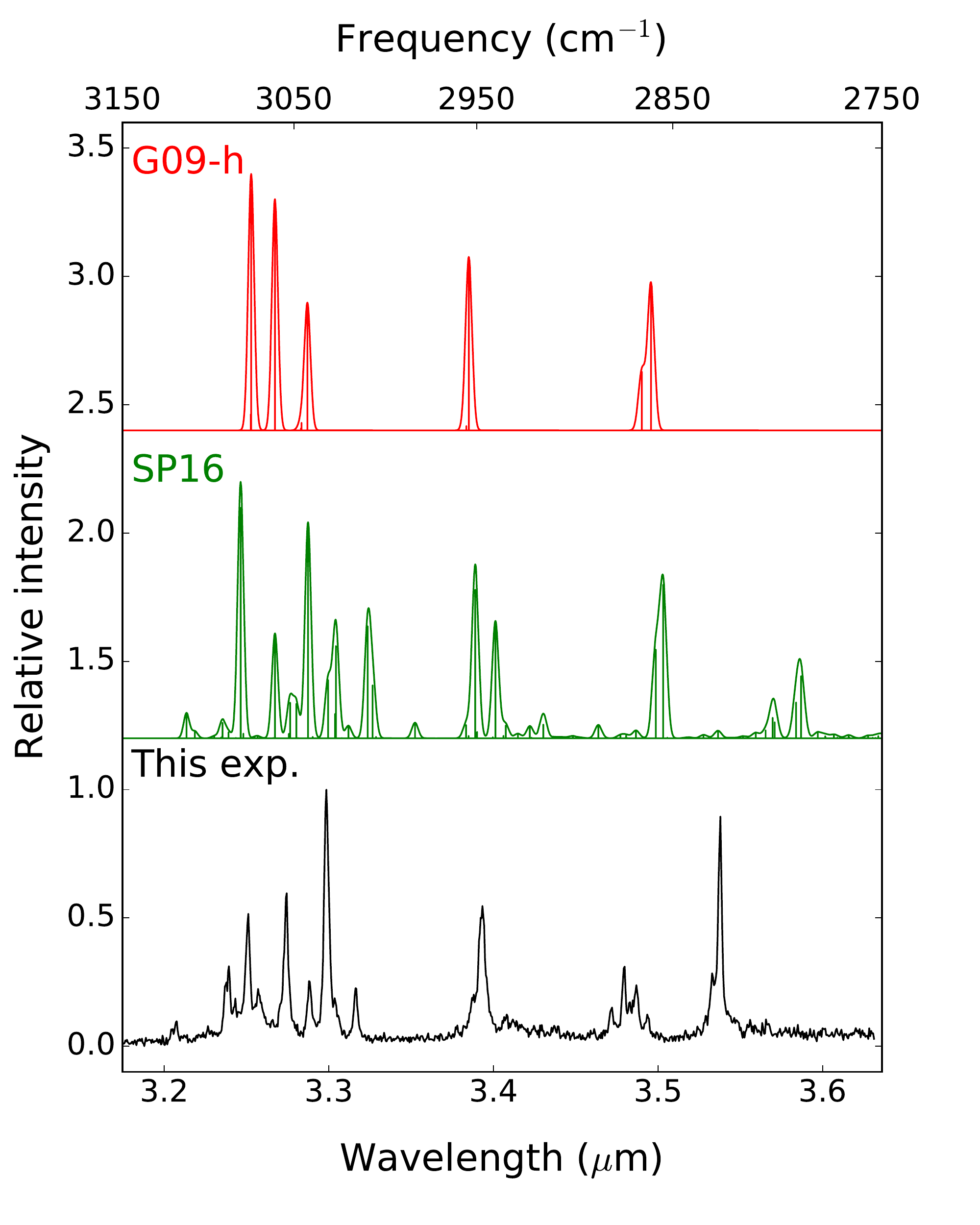} \label{Fig1}}
        \subfigure[DHP]{\includegraphics[scale=0.3]{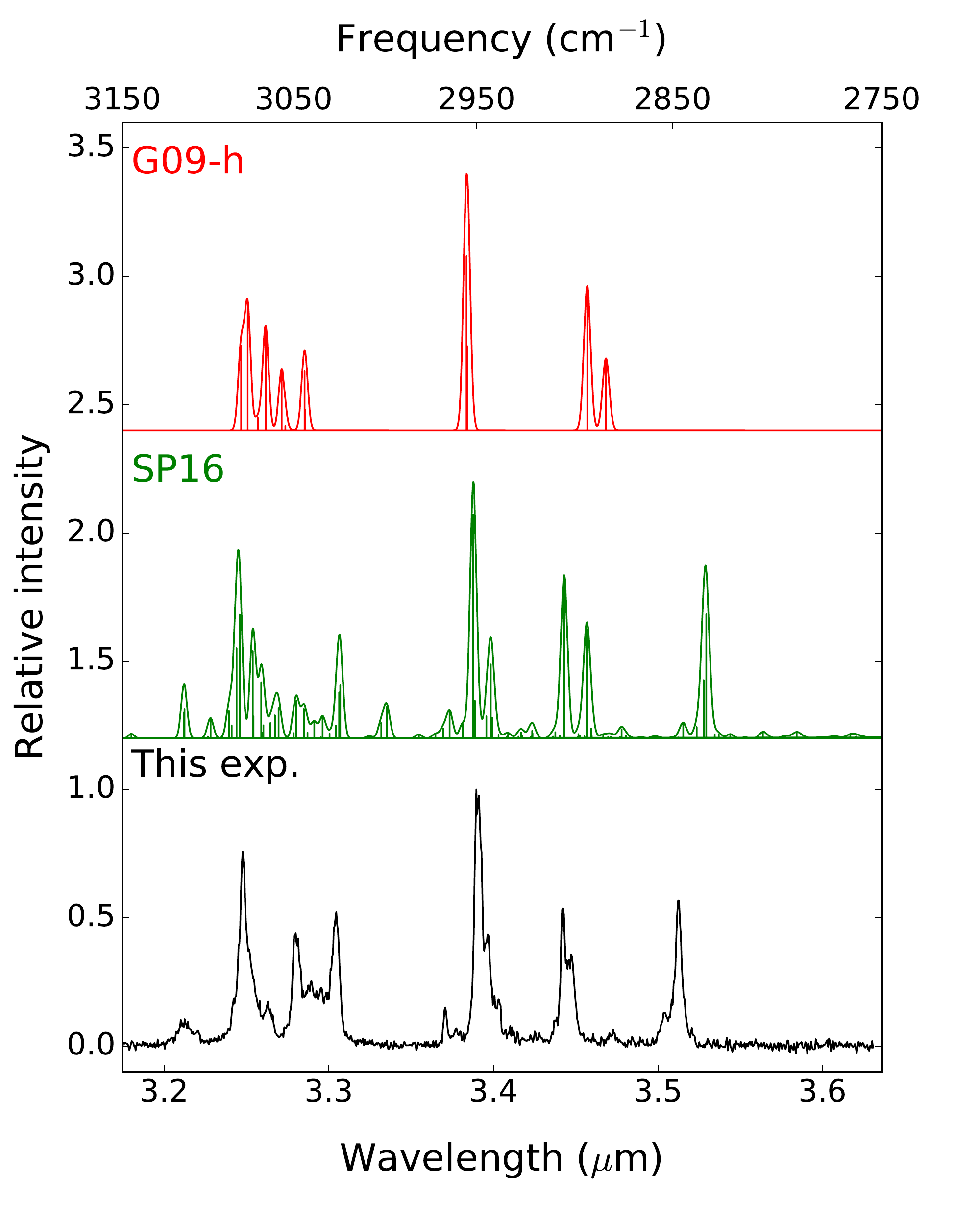} \label{Fig2}}
        \subfigure[THN]{\includegraphics[scale=0.3]{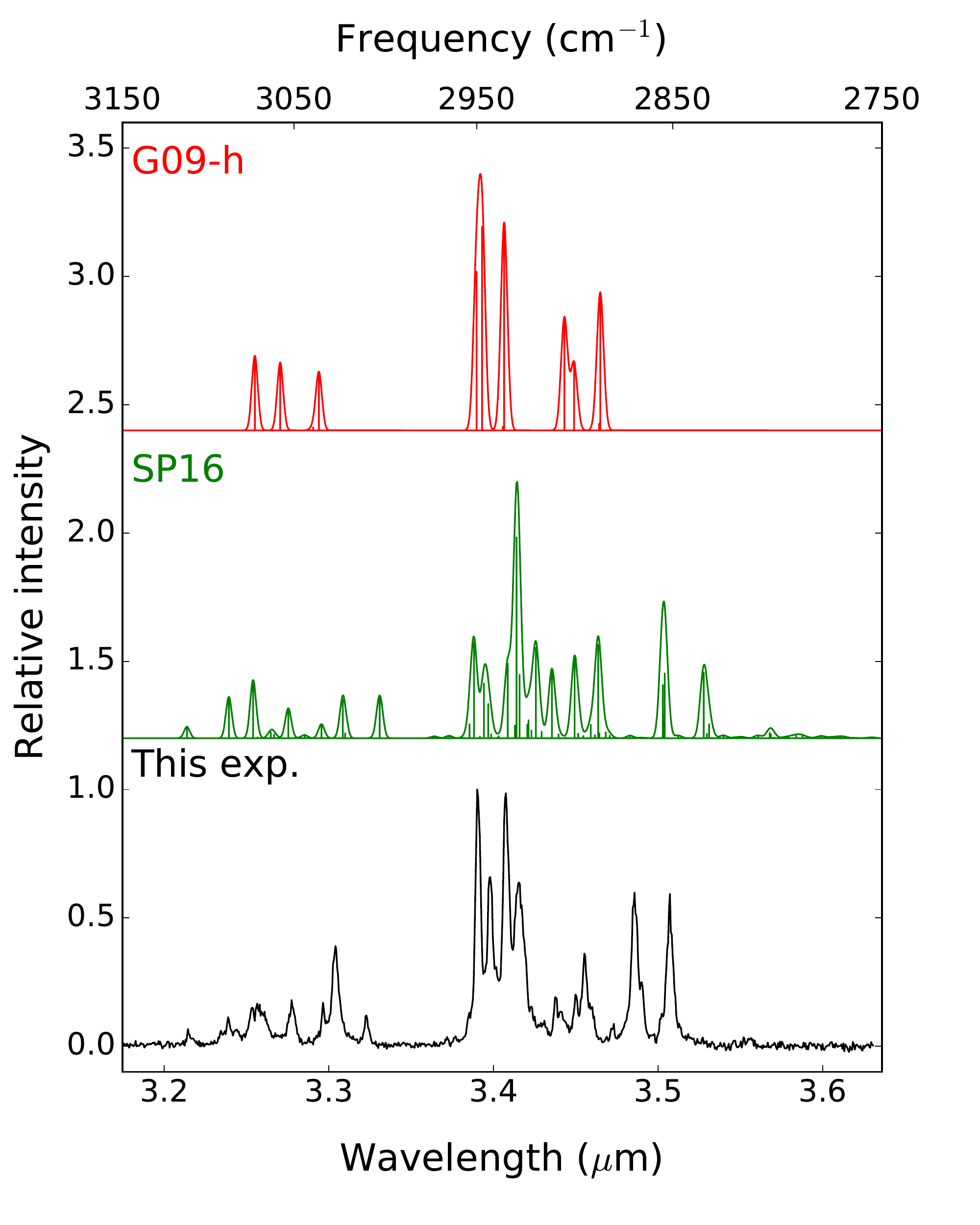} \label{Fig3}}
        \subfigure[HHP]{\includegraphics[scale=0.3]{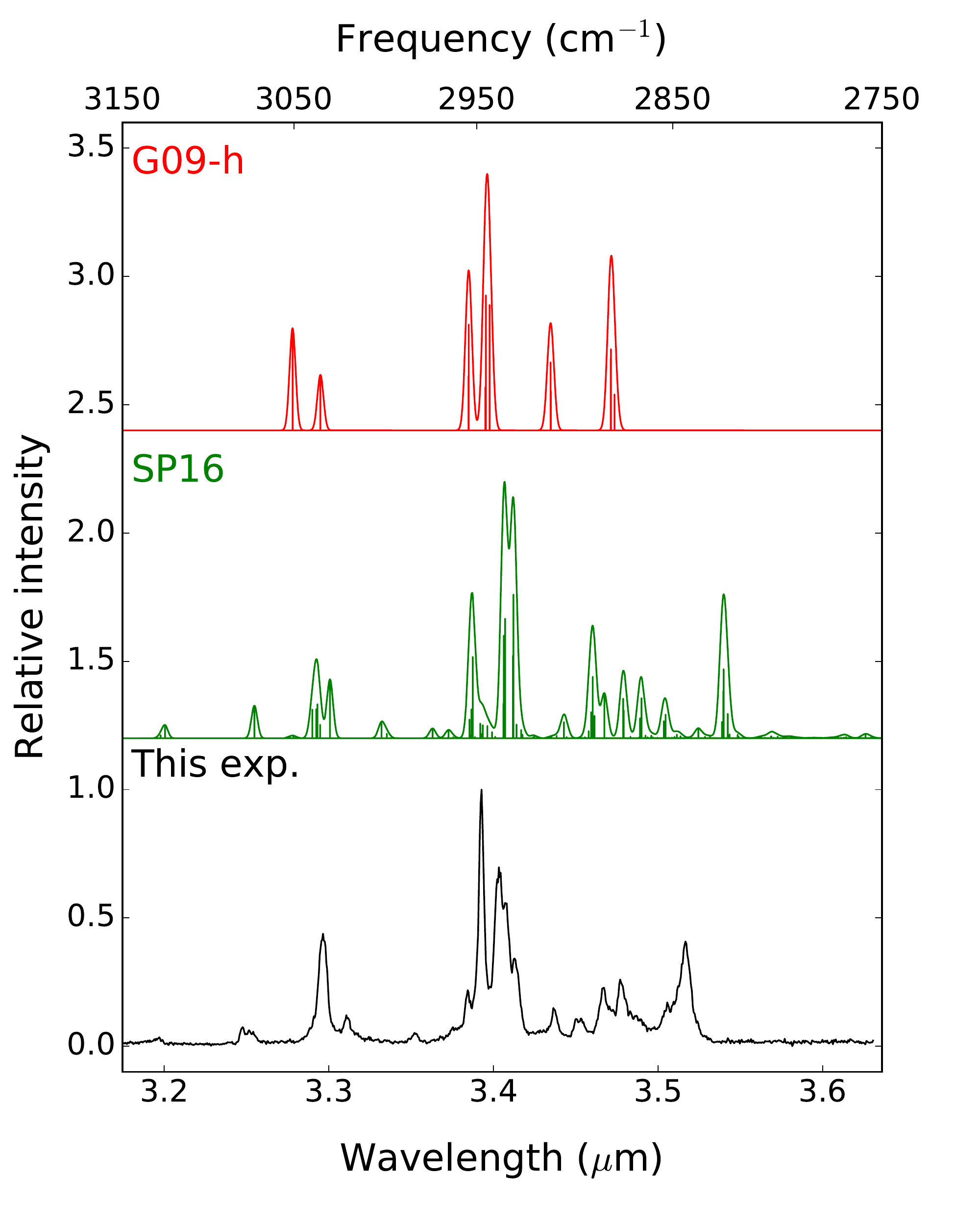} \label{Fig4}}
        \caption{IR absorption spectra of DHA, DHP, THN, and HHP as predicted by G09-h (scaling factor sf= 0.96) and SP16 calculations (not scaled), together with the molecular beam gas-phase spectrum as measured in our experiments. }
\end{figure}

The 3 $\mu$m absorption spectra of jet-cooled DHA, DHP, THN, and HHP in the 3.175-3.636 $\mu$m  (3150-2750 cm$^{-1}$) region are shown in Fig.\ref{Fig1} - \ref{Fig4}, respectively. The  spectra of these molecules display a series of well-separated vibrational bands with line widths ranging from 1 to 7 cm$^{-1}$. The positions and relative line intensities are reported in Table \ref{table1}.

Comparison of these experimental spectra with spectra predicted with the standard harmonic approximation (Fig. \ref{Fig1}-\ref{Fig4}, top panels) appears to give reasonable agreement at first. Closer inspection reveals, however, that the distinct features in the  3.420-3.500 $\mu$m (2924-2857 cm$^{-1}$) region are systematically not predicted. Moreover, considerably more bands are observed in the experimental spectra (Table \ref{table1}) than are expected on the basis of the harmonic approximation. The SP16 anharmonic analysis of these molecules (Fig.\ref{Fig1}-\ref{Fig4}, middle panels), on the other hand, accounts for resonances and predicts more bands to be IR active. In particular, bands observed in the 3.420-3.500 $\mu$m (2924-2857 cm$^{-1}$) region can now be assigned to the overtones and combination bands of CC-stretch and in-plane CH-bending modes that are coupled to the fundamental aliphatic CH-stretch modes through Fermi resonances (detailed assignments are reported in \citep{Mackie2017}.

\onecolumn

\subsection{Methylated PAHs}

\begin{figure}[]
\centering
        \subfigure[MA]{\includegraphics[scale=0.3]{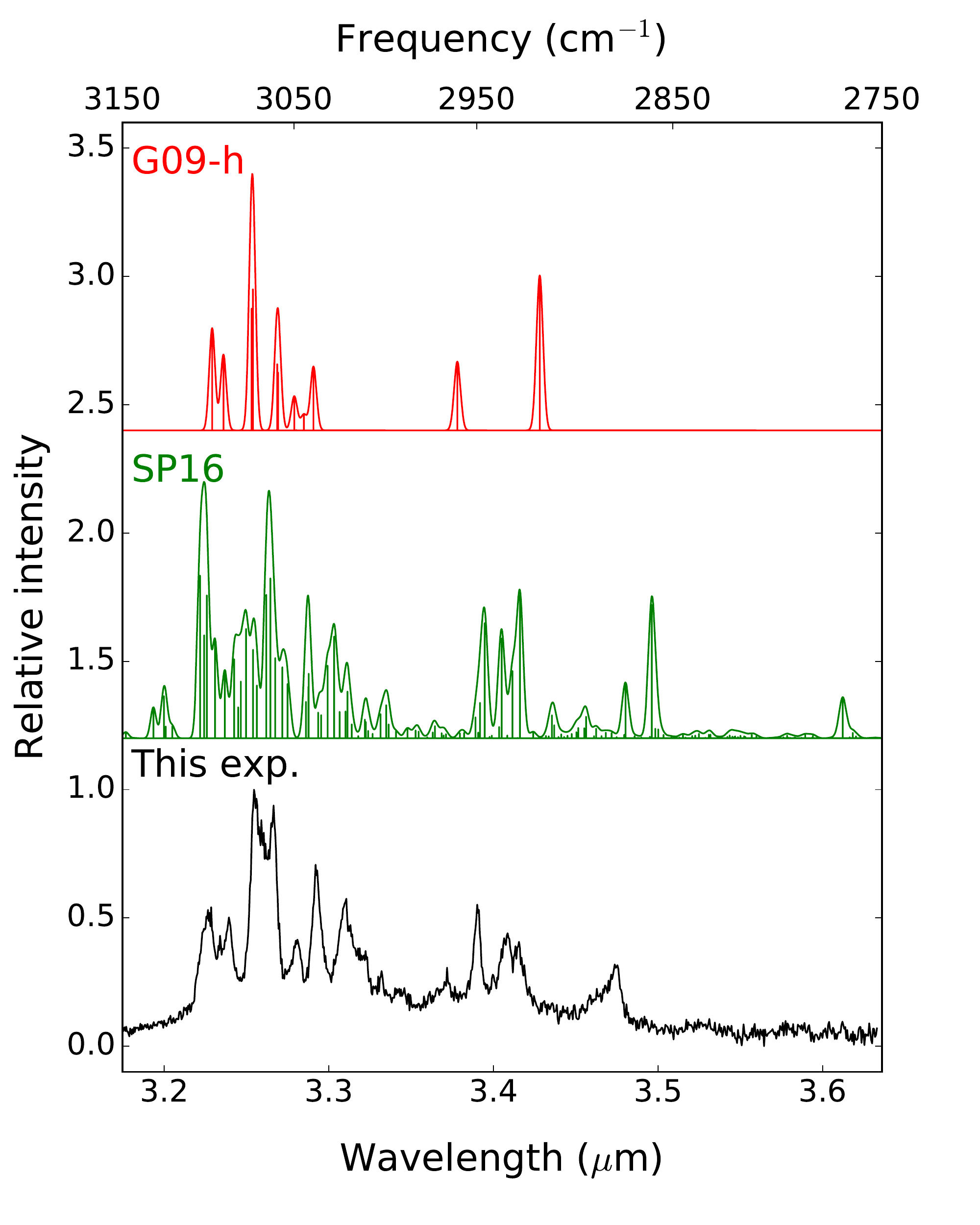}\label{Fig5}}
        \subfigure[DMA]{\includegraphics[scale=0.3]{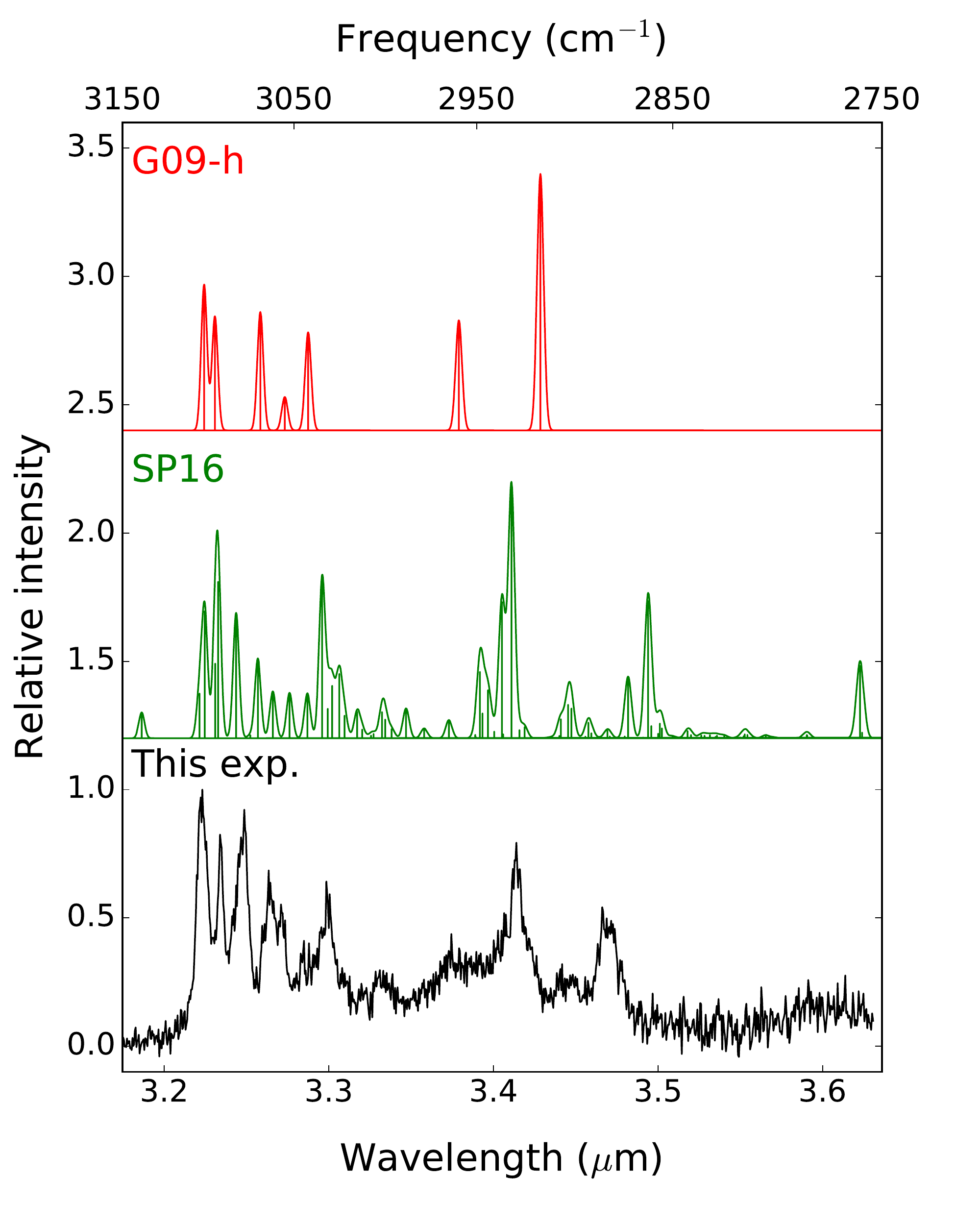}\label{Fig6}}
        \caption{IR absorption spectra of MA and DMA as predicted by G09-h (scaling factor sf= 0.96) and SP16 calculations (not scaled), together with the molecular beam gas-phase spectrum as measured in our experiments.}
\end{figure}

The absorption spectra of jet-cooled 9-methyl\-anthracene (MA) and 9,10-dimethylanthracene (DMA) are depicted in the bottom panels of Fig. \ref{Fig5} and \ref{Fig6}, respectively. These spectra feature broad ($\geqslant$6 cm$^{-1}$) bands that are located on a plateau spanning the 3.215-3.484 $\mu$m (3110-2870 cm$^{-1}$) region (see Table \ref{table1}). In contrast to the experiments, the harmonic approximation predicts only two IR active modes in the alkyl 3.333-3.636 $\mu$m (3000-2750 cm$^{-1}$) region (Fig. \ref{Fig5},\ref{Fig6}, top panels). The SPECTRO VPT2 treatment gives in general a better agreement with the experiment, predicting significantly more bands to be IR active due to Fermi resonances (Figs. \ref{Fig5},\ref{Fig6}, middle panels)). The plethora of bands in the  anharmonic spectra can account for the plateau observed in the 3.215-3.484 $\mu$m (3110-2870 cm$^{-1}$) region of MA and DMA (detailed assignments are reported in \cite{Mackie2017}). 

We find that the anharmonic calculations are in good agreement with the experiment. From an analysis of the harmonic and anharmonic results, we find that it is not so straightforward to separate the aromatic region of the spectrum from the alkyl region as was possible for H-PAHs because of the high-frequency methyl CH-stretch vibrations that occur for wavelengths shorter than 3.333 $\mu$m (3000 cm$^{-1}$).

\section{Discussion}
 
 \begin{figure}[t]
 \centering
        \includegraphics[scale=0.5]{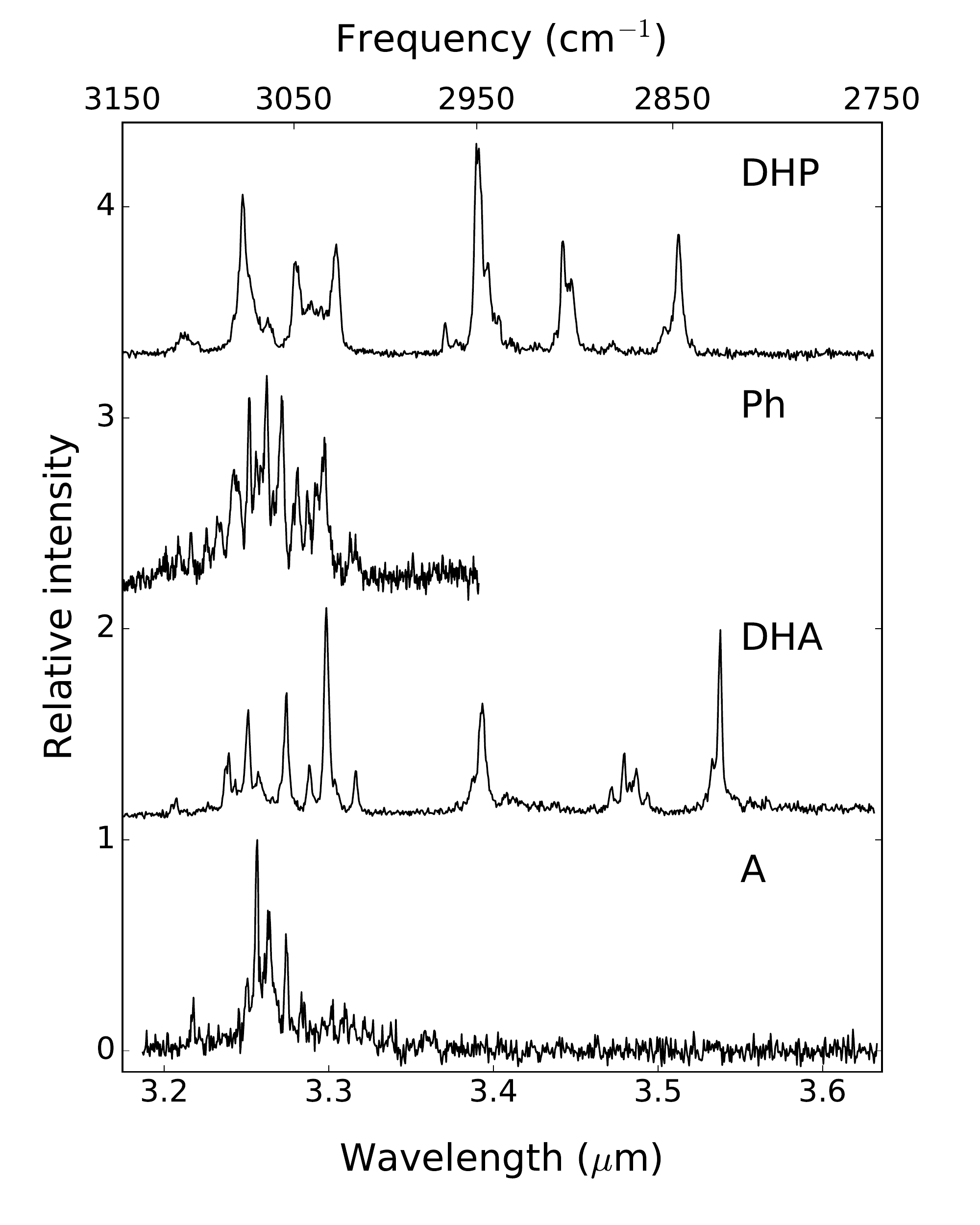}
        \caption{Comparison of IR absorption spectra of jet-cooled A, DHA, Ph, and DHP.  \label{Fig7}}
\end{figure}

\begin{figure}[t]
\centering
        \includegraphics[scale=0.5]{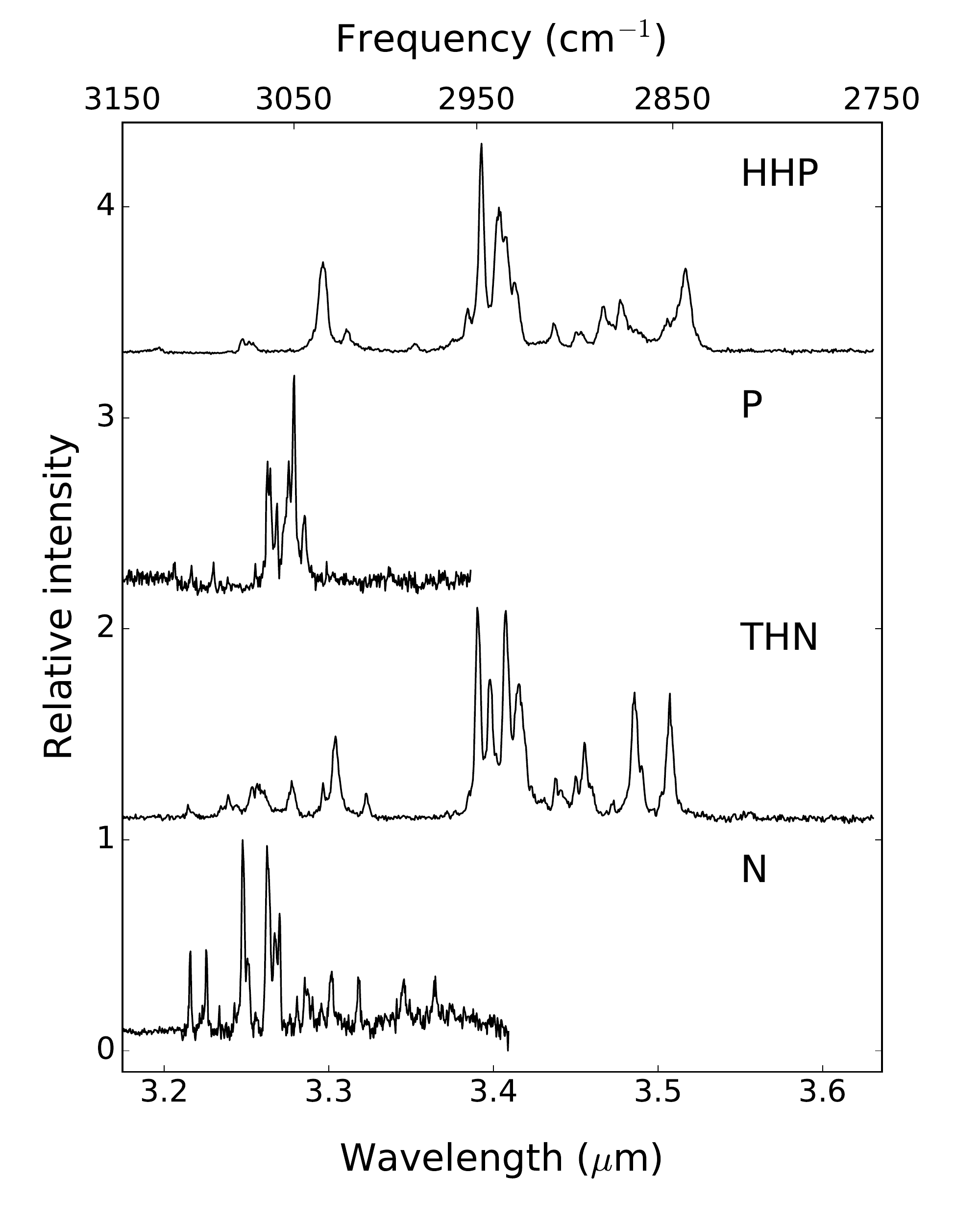}
        \caption{Comparison of IR absorption spectra of jet-cooled N, THN, P, and HHP.  \label{Fig8}}
\end{figure}

\begin{figure}[t]
\centering
        \includegraphics[scale=0.5]{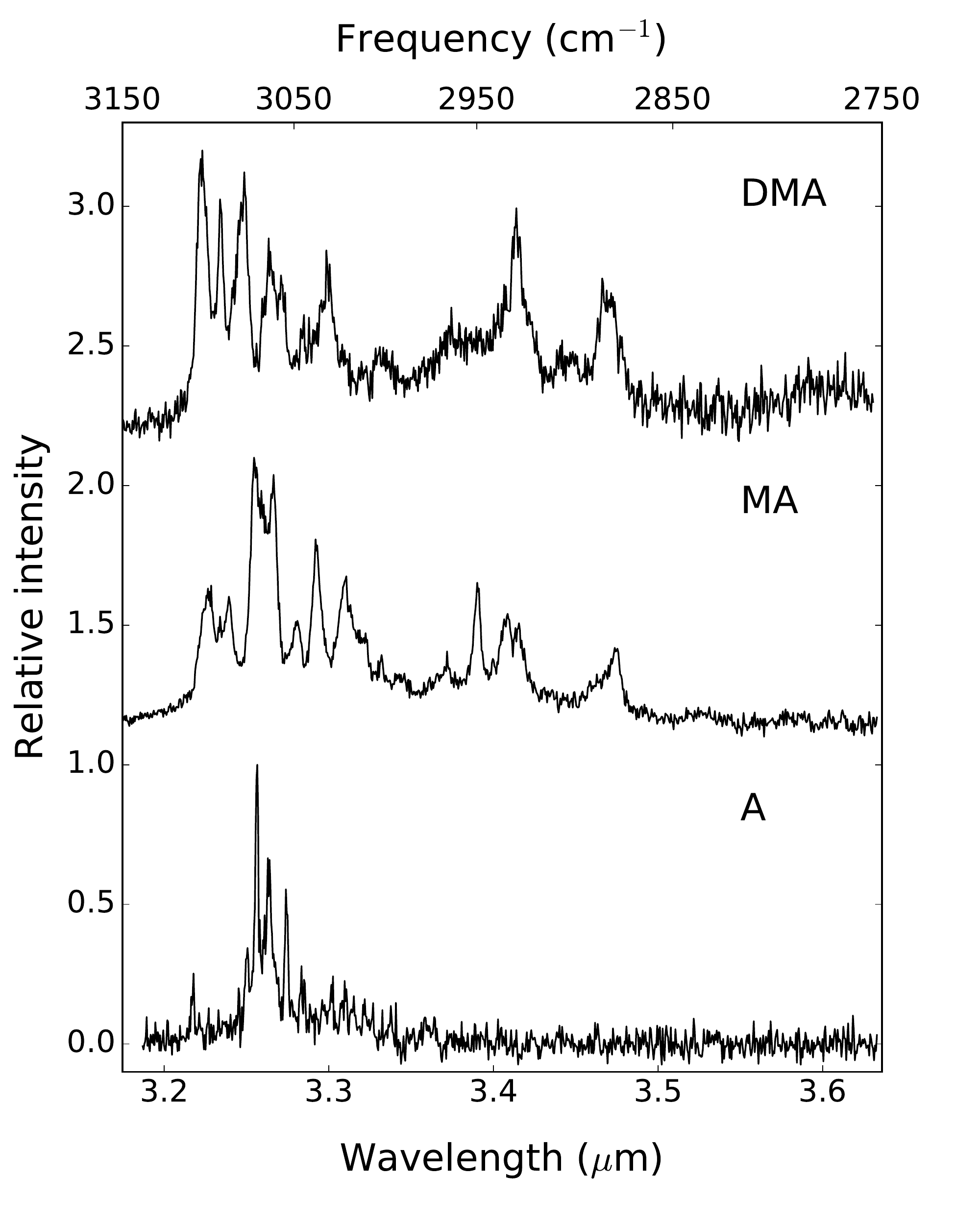}
        \caption{Comparison of IR absorption spectra of jet-cooled A, MA, and DMA.  \label{Fig9}}
\end{figure}

Figures \ref{Fig7}-\ref{Fig9} show a comparison of IR absorption spectra of the decorated PAHs with the spectra of their bare counterpart PAHs, that is, DHA and DHP are compared with anthracene (A) and phenanthrene (Ph); THN and HHP with naphthalene (N) and pyrene (P); and MA and DMA with A. An excess of hydrogens or the presence of a methyl side group have a strong effect on the CH-stretch region of even minimally decorated PAHs like DHA, DHP, and MA. Our study demonstrates that the absorption spectra of these six decorated PAHs in the 3 $\mu$m region can at first approximation be considered as being built up from two (in the
case of H-PAHs, two independent) regions. Their 3.333-3.636 $\mu$m (3000-2750 cm$^{-1}$) region is dominated by the fundamental CH-stretches in the aliphatic or alkyl moieties as well as combination bands and overtones that are coupled with these modes. The 3.175-3.333 $\mu$m (3150-3000 cm$^{-1}$) region, on the other hand, is well known as an aromatic CH-stretch region, although our studies clearly show that this region acquires new properties compared to the CH-stretch region of bare aromatic analogs. In order to understand the spectral changes that the additional hydrogens or methyl-groups promote, several considerations need to be kept in mind.

First, in the case of H-PAHs, the presence of additional hydrogens converts a flat aromatic molecule into a bent, mixed aromatic-aliphatic structure with a lower symmetry (for example, D$_{2h}$ for A vs. C$_{2v}$ for DHA). Similarly, the addition of methyl groups introduces asymmetry as some of its hydrogen atoms do not lie in the molecular plane (D$_{2h}$ for A vs. C$_{s}$ for MA). Such a change in molecular symmetry, along with a larger number of CH-oscillators, increases the total number of normal modes, the number of modes that on symmetry grounds can be IR active, and the number of combination bands that can participate in Fermi resonances. 

Second, attachment of additional hydrogens or methyl groups changes the periphery of the parent molecule. In our previous studies we described the close relation between the shape of the 3.29 $\mu$m band and the edge structure of PAHs \citep{Maltseva2016}. In the case of H- and Me-PAHs, we also observe this tendency, but it is harder to follow because of the increased complexity of the molecular system. For example, absent solo hydrogens in DMA should result in a weaker activity in the low-energy aromatic region around 3.311 $\mu$m (3050 cm$^{-1}$), but the effect is not visible owing to the presence of the sterically hindered methyl stretch at 3.298 $\mu$m (3031.7 cm$^{-1}$; Fig. \ref{Fig9}). The steric interactions between the methyl hydrogens and the quartet hydrogens on the ring also strongly affect the high-energy side of the aromatic region around 3.226 $\mu$m (3100 cm$^{-1}$) and lead to a frequency that is higher than typically observed for these CH-stretches 3.223 $\mu$m (3102.6 cm$^{-1}$). The presence of the methyl group thus not only changes the number of different types of hydrogens (in this case, removing the solo CH-stretch bands from the 3 $\mu$m region), but also adds new methyl CH-stretch modes and introduces high-frequency modes similar to the aromatic CH-stretch modes localized in bay areas of bare PAHs. 

Third, for H-PAHs, the changes in the periphery of the molecules affect the aromatic CH-stretches in a similar way as discussed above for Me-PAHs. In DHA, for example, the only aromatic hydrogens are quartet hydrogens since solo positions are hydrogenated. Since solo CH-stretches typically have the lowest frequencies among the various types of aromatic CH-stretches, we thus expect that hydrogenation of A causes at most a slight blueshift of the absorption profile of the aromatic CH-stretch band. We observe, in contrast, a redshift of the aromatic band (Fig. \ref{Fig7}). The harmonic analysis of the fundamental aromatic CH-stretch modes does not show a significant redshift of the pertinent modes and can thus not account for the observed shift. We therefore conclude that the shift originates from the redistribution of the intensity of the fundamental bands in this region over the combination bands through Fermi resonances. 

Fourth, in addition to the prominent 3.29 $\mu$m band attributed to the aromatic CH-stretch vibrations discussed above, we also observe a series of features in the low-frequency region that are associated with the C(sp$^{3}$)-H stretch vibrations. The absorption spectra show that these features are very similar as long as the PAHs are from the same class (H-PAHs or Me-PAHs). For example, the absorption spectra of hydrogenated species show several well-separated bands restricted to the 3.333-3.636 $\mu$m (3000-2750 cm$^{-1}$) region (Figs. \ref{Fig7} and \ref{Fig8}). Careful assignment of these bands on the basis of the anharmonic calculations reveals that the canonical asymmetric aliphatic CH-stretch, which is assumed to be responsible for the interstellar 3.4 $\mu$m band, is the strongest in three of the four H-PAHs studied here and very similar in terms of frequency (within 4 cm$^{-1}$) for all four molecules (DHA 3.393 $\mu$m (2947 cm$^{-1}$), DHP 3.390 $\mu$m (2950.2 cm$^{-1}$), THN 3.390 $\mu$m (2949.8 cm$^{-1}$), and HHP 3.393 $\mu$m (2947.5 cm$^{-1}$)). The canonical symmetric aliphatic CH-stretch band, in contrast, is more sensitive to the molecular structure. It shows large variations in intensity and frequencies that are distributed over a much wider range  (DHA 3.538 $\mu$m (2826.6 cm$^{-1}$), DHP 3.512 $\mu$m (2847.2 cm$^{-1}$), HHP 3.517 $\mu$m (2843.7 cm$^{-1}$), and THN 3.507 $\mu$m (2851.3 cm$^{-1}$)). For Me-PAHs, we find that MA and DMA show comparable absorption patterns in the 3.333-3.636 $\mu$m (3000-2750 cm$^{-1}$) region (Fig. \ref{Fig9}). The symmetric alkyl CH-stretch gives rise to strong bands at 3.475 $\mu$m (2877.5 cm$^{-1}$) for MA and 3.466 $\mu$m (2885.2 cm$^{-1}$) for DMA, while the asymmetric alkyl CH-stretch is found at 3.409 $\mu$m (2933.8 cm$^{-1}$) and 3.414 $\mu$m (2929.2 cm$^{-1}$) for MA and DMA, respectively, as part of a group of bands in the 3.428-3.356 $\mu$m (2917-2980 cm$^{-1}$) region. 
 
\begin{figure}[t]
\centering
        \includegraphics[scale=1]{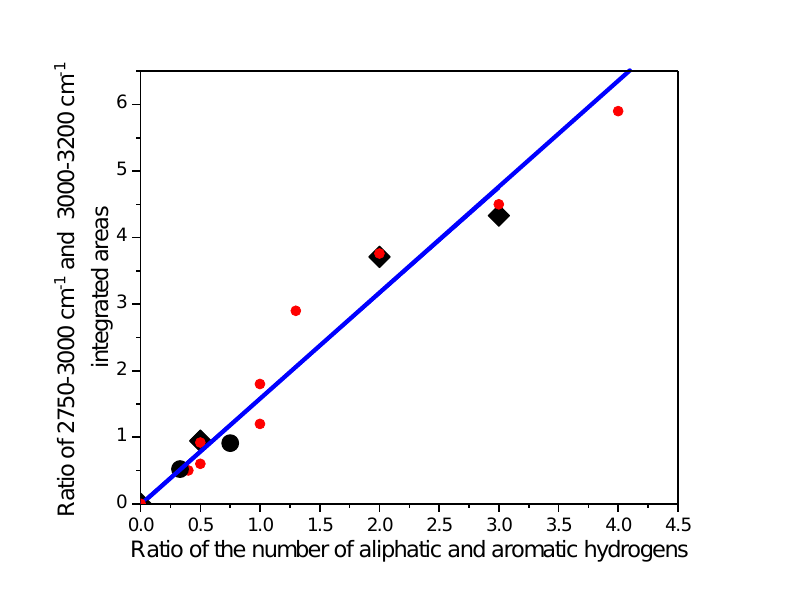}
        \caption{Ratio of integrated intensities of the aliphatic (alkyl) and aromatic CH-stretch regions obtained from our experiments (black diamonds for H-PAHs, black circles for Me-PAHs) and from the NIST database (red circles) as a function of the ratio of the number of aliphatic (alkyl) and aromatic bonds. The blue line is a fit to all data. \label{ratio}}
\end{figure}

Comparison of the absorption spectra of the heavily hydrogenated HHP and THN with the minimally hydrogenated DHP and DHA leads to the conclusion that hydrogenation causes a rapid increase of the integrated band intensities in the aliphatic region and a decrease in the absorption of aromatic bands. Quantitatively, this is expressed in Fig. \ref{ratio}, where the ratio of the integrated intensities of the aliphatic and aromatic CH-stretch bands is plotted as a function of hydrogenation (the ratio of the number of aliphatic and aromatic bonds) for our set of H-PAHs (black diamonds) together with a set of hot gas-phase spectra of 1,3- and 1,4-cyclohexadiene, cyclohexene, 5,12-dihydronapthacene, 1,2,3,4-tetrahydroanthracene, DHA, DHP, HHP, THN from the NIST catalog \citep{NIST} (red circles). We note that our data are completely in line with the independently obtained NIST data, which have been measured under completely different conditions. This demonstrates that our experimental approach is not the reason for the differences with the results from MIS studies (see below). 
The observed linear relation (Fig. \ref{ratio}, blue) is in line with our expectations. Fitting of all data leads to a slope of 1.57$\pm$0.06, which represents the ratio $\alpha$ of aliphatic and aromatic CH-stretch oscillator strengths per CH-bond. This value is in excellent agreement with the value of 1.69 reported in a recent theoretical study \citep{Yang2017}. 

In view of this good agreement, it is surprising that MIS studies of H-PAHs \citep{Sandford2013b} reported a value for $\alpha$ of 2.76, which is significantly higher than concluded here from the gas-phase studies. Further consideration of these results indicates, however, that the difference can be traced back to the influence of the environment on band intensities. It has been shown that incorporation of PAHs into rare gas matrices causes a suppression of intensities of IR bands compared to the isolated molecules \citep{Joblin1994}. Studies like this have not been performed for PAHs with aliphatic CH-moieties, but the observation that $\alpha$ obtained in the MIS studies exceeds $\alpha$ observed in gas-phase experiments suggests that aromatic CH-stretch bands are suppressed to a larger extent under rare gas matrix conditions than aliphatic CH-stretch bands. In view of the fact that a large fraction of the spectroscopic data used in astronomical databases have been acquired under MIS conditions, it is clear that further detailed studies on the influence of the environment on band intensities are important to properly incorporate these data.

The 3 $\mu$m absorption of methylated anthracenes differs significantly in a number of aspects from its hydrogenated analogs (Figs. \ref{Fig7} and \ref{Fig8} vs. Fig. \ref{Fig9}). First of all, unlike H-PAHs, the low-energy side of the aromatic CH-stretch band of Me-PAHs has a contribution from the CH-stretch bands of the methyl group. Second, the absorption attributed to the alkyl group starts at 3.484 $\mu$m (2870 cm$^{-1}$), which is 60 cm$^{-1}$ higher than the absorption of hydrogenated PAHs. Third, the absorption spectra of methylated PAHs display broad and poorly separated bands, but even more strikingly, a prominent plateau that covers the entire 3.215-3.484 $\mu$m (3110-2870 cm$^{-1}$) region. This plateau is a consequence of the high density of accessible states, which in turn is caused by the low symmetry and the increased number of normal modes, as is confirmed by the anharmonic calculations that show a wide variety of resonances. The plateau can also be related to the nearly ''free rotor" nature of the methyl groups for which the theory cannot account at present \citep{Mackie2017}. 

\begin{figure}[t]
\centering
        \includegraphics[scale=0.5]{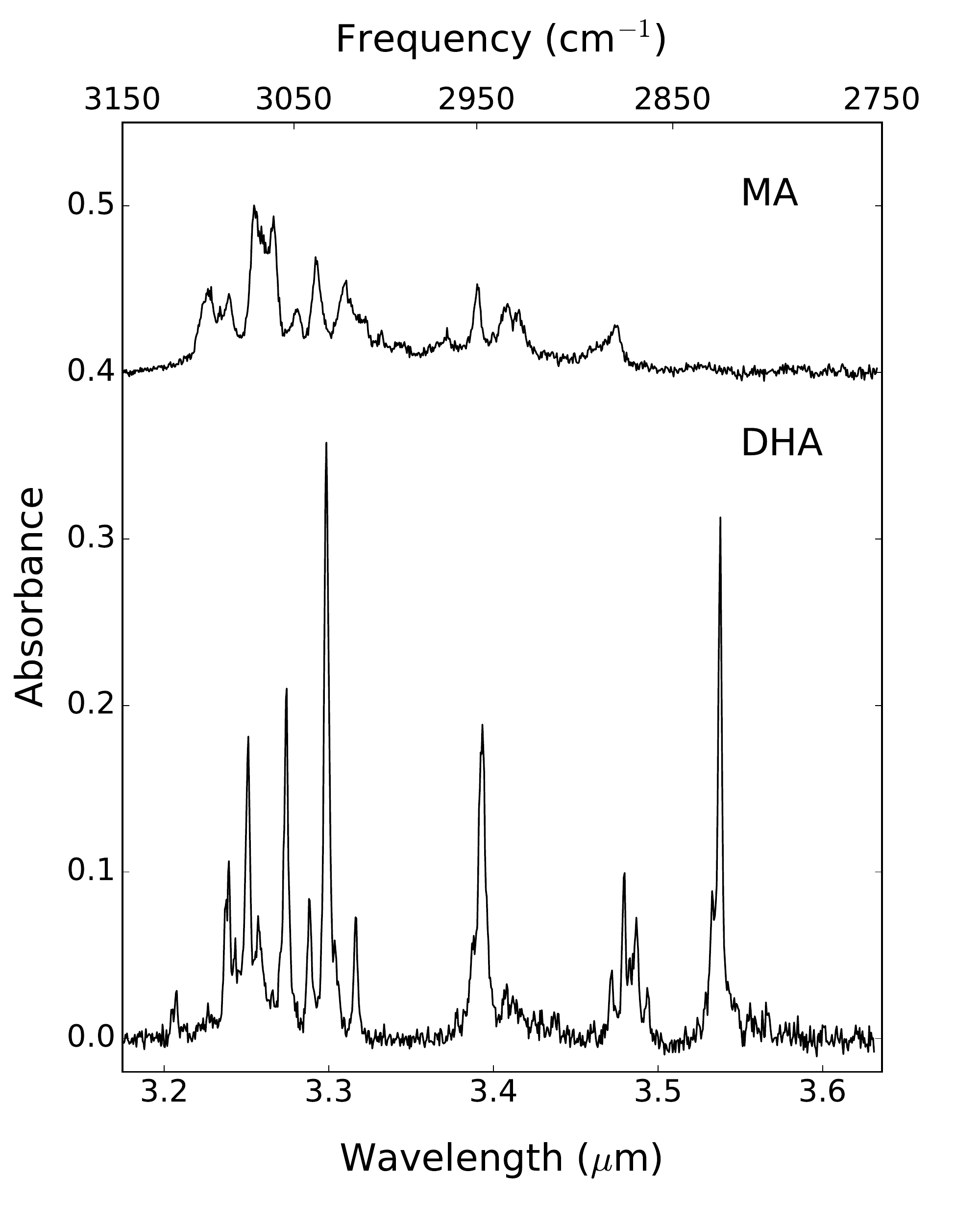}
        \caption{Comparison of the experimental IR absorption spectra of jet-cooled DHA and MA. \label{DHAMA}} 
\end{figure}    

Figure \ref{DHAMA} shows a comparison of the absorbances of DHA and MA instead of the relative scale shown in Figs. \ref{Fig1} and \ref{Fig5}. A remarkable observation is the difference in the maximum absorption intensity in the two spectra. At the same time, we find that the absorption intensity integrated over the entire spectrum  as well as the ratio of integrated intensities of alkyl and aromatic CH-stretches (Fig. \ref{ratio}, black circles) is quite similar, and this is indeed what is also predicted by the calculations \citep{Yang2017}. From this comparison we thus conclude that the significantly reduced peak intensities of fundamental transitions in the 3 $\mu$m region of MA as compared to DHA are due to a redistribution of the intensity over many more combination bands. This is important as it indicates that for alkylated PAHs, peak intensities can in general be expected to be lower than for H-PAHs. 
 
\section{Astrophysical implications}

We here extended our studies of PAHs into the aliphatic and alkyl domains, demonstrating that non-aromatic CH-stretch modes actively participate in Fermi resonances, similar to what has previously
been concluded for aromatic CH-stretches. It is thus clear that anharmonicity and resonances define the shape and the strength of the absorption bands in the 3 $\mu$m region and cannot be neglected. 

The experimental spectra presented here suggest that the carriers of the 3.4 $\mu$m emission band are H-PAHs. We find that the aliphatic asymmetric CH-stretch modes in hydrogenated PAHs (i) are not sensitive to the molecular structure and thereby appear in a restricted frequency range (3.389-3.393 $\mu$m (2947.0-2950.2 cm$^{-1}$)), and (ii) they are among the strongest bands as compared to other CH-stretch bands. The asymmetric alkyl CH-stretch of MA and DMA (3.414-3.409 $\mu$m (2929-2933 cm$^{-1}$)) also seems to indicate that it is insensitive to structural details. However, in view of the band intensities and bearing in mind that in emission, a redshift occurs, the aliphatic asymmetric CH-stretch is more likely to contribute to the 3.4 $\mu$m band than the methyl CH-stretch in Me-PAHs. The latter can, however, be responsible for the low-energy shoulder of the 3.4 $\mu$m band that is observed, for example, in IRAS 21282-5050 \citep{JourdaindeMuizon1990}. Our conclusion on the carriers of the 3.4 $\mu$m band, however, contradicts with recent theoretical study of \cite{Andrews2016}. These experimental results supported by previous studies \citep{Wagner2000a} demonstrate that H-PAHs provide a better match to the observed position of the main 3.4 $\mu$m band than Me-PAHs and thus suggest that  our understanding of the hydrogenation behavior of PAHs in photon-dominated regions is still incomplete.

The symmetric aliphatic CH-stretch shows a greater dependence on structural details. Here we find a variation of 25 cm$^{-1}$ in band position (3.538-3.507 $\mu$m (2826-2851 cm$^{-1}$)). It might very well be that the astronomically observed 3.51 $\mu$m band is due to the overlap of this band for different PAHs. This conclusion is supported by the observation of the Orion bar, where it was established that the 3.40 and 3.51 $\mu$m bands share a common origin \citep{Sloan1997,Geballe1989a}. 

Energetically, the lowest CH-stretch bands of MA and DMA are found at 3.475 $\mu$m (2877.5 cm$^{-1}$) and 3.466 $\mu$m (2885.2 cm$^{-1}$). It is tempting to correlate this band with the 3.46 $\mu$m emission band, although the blueshift of the MA and DMA data cannot be explained yet. Further high-resolution spectroscopic studies of a more extensive set of methylated PAHs are important
in this respect as they will provide further support for a correlation with the 3.46 $\mu$m band. It is well known that the 3.46 $\mu$m emission band is more prominent in carbon-rich sources such as the planetary nebula IRAS 21282+5050 \citep{JourdaindeMuizon1986} than in the Orion bar or other star-forming regions \citep{Sloan1997}. Our data thus suggest that the carrier composition in carbon-rich regions contains a significant fraction of methylated PAHs along with hydrogenated ones, while no such abundance of methylated PAHs is expected for ionization regions. At the same time, it must also be noted that in the Orion bar, no correlation has been found between an extra component of the 3.40 $\mu$m band and the 3.46 $\mu$m band \citep{Sloan1997}. It would be useful to investigate this correlation for other sources such as protoplanetary nebulae, where these features are more prominent. Another feature observed in a wide range of sources is a band at 3.56 $\mu$m. As yet, no conclusive assignment has been made for this band. Our studies, however, suggest that it does not derive from H- and Me-PAHs.
 
On the basis of our present and previous studies, we conclude, supporting previous IR studies \citep{Wagner2000a}, that the plateau extending in the 3.2-3.6 $\mu$m region likely has a complex origin to which all studied subclasses of molecules are able to contribute. The 3.17-3.33 $\mu$m region can originate from the overlap of combination bands and overtones coupled to aromatic CH-stretches in bare PAHs, H- and Me-PAHs, and small contributions from the coupling to hindered methyl CH-stretch vibrations. The region 3.33-3.64 $\mu$m is fully defined by the IR activity in the aliphatic and alkyl regions of H- and Me-PAHs, although the way they contribute is different. H-PAHs show a series of distinctive bands, while Me-PAHs display a broad absorption plateau with on top low-intensity primary bands.

The ratio of the intrinsic strengths of the aliphatic and aromatic CH-stretch vibrations ($\alpha$=1.58) obtained in this work allows for a more accurate estimate of the aliphatic fraction. Assuming that the the 3 $\mu$m plateau and the satellite features on top of it derive from H-PAHs, we estimate that for IRAS 21282+5050 and NGC 1333, IR sources with a normal 3 $\mu$m profile, the carriers of the 3 $\mu$m emission contain two to three aliphatic hydrogens per eight aromatic hydrogens, while for IR emitters with an anomalously strong 3.4 $\mu$m feature, the fraction of aliphatic hydrogens is much higher. For IRAS 22272+5435, six to seven aliphatic per eight aromatic hydrogens are found, while for IRAS 04296+3429, the aliphatic hydrogens even outnumber the aromatic hydrogens (more than nine aliphatic per eight aromatic hydrogens). These estimates contradict recent attempts to quantify the aliphatic fraction by means of MIS studies \citep{Sandford2013b}, in which it was concluded that the aromatic moieties exceed aliphatic ones even in anomalous objects like IRAS 04296+3429. On the other hand, our results corroborate previous gas-phase studies on methyl-coronene, where for PAHs in NGC 1333, one methyl side group per eight aromatic hydrogens was estimated \citep{Joblin1996}.  

Finally, we emphasize that our studies show that not only the 3.33-3.64 $\mu$m region undergoes significant changes upon hydrogenation and/or methylation, but also the region of the aromatic CH-stretch vibrations (3.17-3.33 $\mu$m). This has to be taken into account in the analysis of the 3.29 $\mu$m band, especially for sources with a significant intensity of the 3.33-3.64 $\mu$m region, as hydrogenated and methylated species are expected to be more present in these sources and contribute to the aromatic region.

\section{Conclusions}

We have presented high-resolution IR spectra of six jet-cooled decorated PAHs in the 3 $\mu$m region. The results of our study are in line with the conclusions from our previous papers on acenes and condensed PAHs and clearly show that anharmonicity plays an important role in the CH-stretch region of hydrogenated and, in particular, methylated PAHs. In combination with the high density of states, this results in a multitude of bands that acquire intensity through Fermi resonances. While for bare PAHs we concluded that anharmonicity can only account for part of the 3 $\mu$m plateau, decorated PAHs give rise to plateaus that extend up to over 3.6 $\mu$m.

Our studies demonstrate that the 3.33-3.64 $\mu$m region of hydrogenated and methylated PAHs shows a substantial fraction of intensity that can easily exceed the fraction of intensity associated with the aromatic region. It is therefore very likely that these species are carriers of the low-frequency features in the 3 $\mu$m region observed by astronomers. We have indeed found that the fundamental symmetric and asymmetric CH-stretches of the methylene part of  H-PAHs and of the methyl group of Me-PAHs are remarkably consistent with the 3.4 and 3.51 and with the 3.41 and 3.46 $\mu$m bands in UIR emission. In order to distinguish between these subclasses, high-resolution studies of the methylene scissoring region (6.9 $\mu$m) and the periphery-sensitive CH out-of-plane region (10-15 $\mu$m) would be very worthwhile. Such studies are currently in preparation.  

\begin{acknowledgements}
The experimental work was supported by The Netherlands Organization for Scientific Research (NWO). This work has been performed as part of the Dutch Astrochemistry Network (DAN). AP acknowledges NWO for a VIDI grant (723.014.007). Studies of interstellar PAHs at Leiden Observatory have been supported through a Spinoza award. PAH studies at Leiden and Nijmegen are supported through the Marie Skodowska-Curie initial training network, EUROPAH under the Horizon2020 framework. The theoretical work was carried out on the Dutch national e-infrastructure with the support of SURF Cooperative (project SH-362-15). AC acknowledges NWO for a VENI grant (639.041.543). XH and TJL gratefully acknowledge support from the NASA 12-APRA12-0107 grant. XH acknowledges the support from NASA/SETI Co-op Agreement NNX15AF45A. Some of this material is based upon work supported by the National Aeronautics and Space Administration through the NASA Astrobiology Institute under Cooperative Agreement Notice NNH13ZDA017C issued through the Science Mission Directorate
\end{acknowledgements}

\onecolumn

\begin{table}
\caption{Experimental line position (cm$^{-1}$) and relative intensities for the absorption bands of THN, DHA, DHP, HHP, MA, and DMA in 3.175-3.636 $\mu$m  region.}\label{table1}
\centering
\begin{tabular}{lrlrlrlrlrlr}
\hline\hline

\multicolumn{2}{c}{THN (C$_{10}$H$_{12}$)}&\multicolumn{2}{c}{DHA (C$_{14}$H$_{12}$)}&\multicolumn{2}{c}{DHP (C$_{14}$H$_{12}$)}&\multicolumn{2}{c}{HHP (C$_{16}$H$_{16}$)}&\multicolumn{2}{c}{MA (C$_{15}$H$_{12}$)}&\multicolumn{2}{c}{DMA (C$_{16}$H$_{14}$)}\\

freq.& rel. int.&freq.& rel. int.&freq.& rel. int.&freq.& rel. int.&freq.& rel. int.&freq.& rel. int. \\
\hline
2851.3 & 0.6  & 2826.6 & 0.9  & 2847.2 & 0.57 & 2843.7 & 0.4  & 2877.5 & 0.32 & 2885.2 & 0.54 \\
2865.3 & 0.25 & 2830.6 & 0.28 & 2854.5 & 0.13 & 2852.7 & 0.17 & 2927.9 & 0.4  & 2929.2 & 0.79 \\
2868.9 & 0.6  & 2862.6 & 0.12 & 2879.7 & 0.07 & 2876.1 & 0.26 & 2933.8 & 0.44 & 2963.5 & 0.44 \\
2879.3 & 0.09 & 2868   & 0.24 & 2901   & 0.36 & 2884.3 & 0.2  & 2949.8 & 0.55 & 3031.7 & 0.64 \\
2894.2 & 0.36 & 2871.1 & 0.17 & 2905.1 & 0.54 & 2896   & 0.1  & 2965.8 & 0.3  & 3057.3 & 0.55 \\
2898.8 & 0.2  & 2873.8 & 0.31 & 2938.3 & 0.18 & 2910.1 & 0.15 & 3001.2 & 0.3  & 3078.3 & 0.92 \\
2906.5 & 0.14 & 2880.2 & 0.15 & 2944.3 & 0.44 & 2930.1 & 0.34 & 3020.6 & 0.58 & 3092.3 & 0.83 \\
2909.2 & 0.19 & 2934.2 & 0.12 & 2950.2 & 1    & 2934.7 & 0.56 & 3037.8 & 0.7  & 3102.6 & 1    \\
2927.9 & 0.64 & 2947   & 0.55 & 2966.7 & 0.15 & 2938.3 & 0.7  & 3048   & 0.41 &        &      \\
2934.7 & 0.99 & 2951.6 & 0.2  & 3026.2 & 0.52 & 2947.5 & 1    & 3061.5 & 0.94 &        &      \\
2942.9 & 0.66 & 3015.5 & 0.23 & 3034.5 & 0.23 & 2954.8 & 0.22 & 3068.5 & 0.88 &        &      \\
2949.8 & 1    & 3027.1 & 0.19 & 3040.1 & 0.25 & 2982.3 & 0.05 & 3072.7 & 1    &        &      \\
3009.5 & 0.12 & 3031.7 & 1    & 3048.9 & 0.44 & 3020.6 & 0.12 & 3087.1 & 0.5  &        &      \\
3026.6 & 0.39 & 3041.5 & 0.25 & 3064.7 & 0.18 & 3033.6 & 0.44 & 3092.3 & 0.43 &        &      \\
3033.6 & 0.17 & 3054   & 0.6  & 3079.2 & 0.76 & 3075.5 & 0.06 & 3099.3 & 0.53 &        &      \\
3051.2 & 0.18 & 3070.3 & 0.22 & 3084.3 & 0.19 & 3079.2 & 0.07 &        &      &        &      \\
3066.6 & 0.14 & 3075.9 & 0.52 & 3112.9 & 0.1  & 3128   & 0.04 &        &      &        &      \\
3070.8 & 0.17 & 3083.4 & 0.19 &        &      &        &      &        &      &        &      \\
3073.6 & 0.15 & 3087.1 & 0.31 &        &      &        &      &        &      &        &      \\
3082.5 & 0.07 & 3089.5 & 0.23 &        &      &        &      &        &      &        &      \\
3087.6 & 0.1  & 3099.3 & 0.08 &        &      &        &      &        &      &        &      \\
3111   & 0.07 & 3118.1 & 0.1  &        &      &        &      &        &      &        &                 \\
\hline
\end{tabular}
\end{table}

\end{document}